\documentstyle[12pt]{article}

\begin{document}
\pagestyle{plain}

\title{\bf  Solitons  in the Camassa - Holm \\ Shallow Water Equation}
\author{
Fred Cooper \\
{\small \sl Theoretical Division, Los Alamos National Laboratory,}\\
{\small \sl Los Alamos, NM 87545}\\
{\small \sl and   }\\
{\small \sl Physics Department, University of New Hampshire,} \\
{\small \sl Durham, NH 03824}\\
\and Harvey Shepard \\
{\small \sl Physics Department, University of New Hampshire,} \\
{\small \sl Durham, NH 03824}\\
\\
}
\maketitle

\begin{abstract}
We study the class of shallow water equations of Camassa and Holm
derived from the Lagrangian: \\
$ L= \int \left( \frac{1}{2} (\varphi_{xxx}-\varphi_{x} )\varphi_{t} -  {1
\over 2}
{(\varphi_{x})^{3}}  - {1 \over 2}\varphi_{x}(\varphi_{xx})^{2} - {1 \over 2}
\kappa \varphi_{x}^{2}  \right) dx, $ \\
This class contains ``peakons'' for $\kappa=0$,  which
are solitons  whose peaks have a discontinuous first derivative.  We
derive approximate solitary wave solutions to this class of equations
 using  trial variational functions of the form $u(x,t) = \varphi_{x}= A(t)
\exp
\left[-\beta (t) \left|x-q(t)  \right|^{2n} \right]$  in a time-dependent
variational calculation. For the case $\kappa=0$ we obtain the exact
answer. For $\kappa \neq 0$ we obtain  the optimal variational
 solution. For the variational solution having fixed conserved  momentum $P =
\int
\frac{1}{2} (u^2 +u_x^2)  dx $,  the soliton's scaled amplitude, $A/P^{1/2}$,
and velocity, $\dot{q}/P^{1/2}$, depend only on the variable $z={\kappa \over
\sqrt{P}}$.   We  prove  that these scaling relations are true for the exact
soliton solutions to the Camassa-Holm equation.

\vspace{7 mm}
\vspace{7 mm}

\end{abstract}

\section{Introduction}

Recently, Camassa and  Holm \cite{CH} have shown that a particular
generalization of the Benjamin-Bona-Mahoney (BBM) equation
\cite{BBM} ,
\begin{equation}
  u_t + \kappa u_x -u_{xxt} +3u u_x = 2 u_x  u_{xx }+u u_{xxx} , \label{eq1}
\end{equation}
is completely  integrable for all  $\kappa$. That is,  eq. (\ref{eq1}) supports
solutions on the real line that show the usual elastic collision properties of
solitons.   For the case $ \kappa=0$ the solitons
have peaks with discontinuous first derivatives
(so-called ``peakons'') and for N-interacting peakons, $u$ has
the form
\begin{equation}
u(x,t) = \sum
_{i=1}^N p_i(t) exp(-|x-q_i(t)|),
\end{equation}
where $p$ and $q$ are canonically conjugate variables satisfying
Hamilton's equations with Hamiltonian
\begin{equation}
H= {1 \over 2} \sum_{i,j=1} ^ {N} p_i p_j e^ {- |q_i - q_j | }.
\end{equation}
In \cite{CH}  it was also noted that for $\kappa \neq 0$
the solitons are no longer peaked.

 In previous papers \cite{cs1}  \cite{cs2}  \cite{cs3} we have
    demonstrated the versatility and robustness of the
class of trial wave functions of the form: \\
$u(x,t) =  A(t) \exp
\left[-\beta (t) \left|x-q(t)  \right|^{2n} \right]$ , in that they provide
an excellent  approximation to the true single solitary wave
behavior for a variety of nonlinear equations derivable from an
action principle. In the present  paper we use a variational
approach and obtain the
exact soliton for the $\kappa=0$ case. For the general
case of $\kappa \neq 0$, we identify the relevant
parameter that governs all the properties of the solitary wave.  We find  that
the variational solution  having fixed conserved  momentum $P = \int
\frac{1}{2} (u^2 +u_x^2)  dx, $  depends only on the variable  $z={\kappa \over
\sqrt{P}}$.  We  also show that the velocity of the solitary wave  can be
determined by
  $\dot{q}= P^{1/2} F(z)$ where    $F(0)=1$. For $z \neq 1$, $F(z)$ can
be determined analytically for small z and  numerically otherwise.  The
variational calculation also leads to the result that the parameters $A$ and
$\beta$ describing the variational soliton wave function are time independent
and depend only on the parameter $z$.

The Lagrangian for the Camassa-Holm equation \cite{ch2} is:
\begin{equation}
L= \int \left( \frac{1}{2} (\varphi_{xxx}-\varphi_{x} )\varphi_{t} - {1 \over
2}
\varphi_{x}^{3} - {1 \over 2}\varphi_{x} \varphi_{xx}^{2} - {\kappa \over 2}
\varphi_{x}^{2}  \right) dx.  \label{L}
\end{equation}
This identifies the conserved Hamiltonian to be ($ u=\varphi_x $):
\begin{equation}
H=  \int  dx \left({1 \over 3}  u^{3}+ {1 \over 2} u u_{x} ^2 + {1 \over
2} \kappa u ^{2}  \right) .\label{H}
\end{equation}
{}From the translational invariance of the Lagrangian (\ref{L}),
 Noether's theorem implies the conservation of the momentum $P$:
\begin{equation}
P= \frac{1}{2} \int  dx \left( u^{2}+u_{x}^{2}  \right) .\label{P}
\end{equation}

\section{Variational Method}

If we insert our variational Ansatz for the solitary wave,
\begin{equation}
u(x,t) =  A(t) \exp
\left[-\beta (t) \left| x-q(t) \right|^{2n}  \right], \label{var1}
\end{equation}
into the Lagrangian (\ref{L}) and use (\ref{P}) to eliminate $A$ in
favor of $P$, we obtain
\begin{equation}
L= P \dot{q} - H( P,n,\rho) \label{l2}
\end{equation}
where $\rho \equiv \beta^{-1/2n}$.
We see that $P$ is conjugate to $q$ in the sense that it generates the spatial
translations. The collective coordinate Hamiltonian
$H( P,n,\rho) $  has $\beta (\rho)$ as a cyclic variable and thus
$\rho$ is time independent. $H( P,n,\rho) $  is also independent of
$q$  so that the canonical momentum P is still conserved for the variational
 wave function.  Since  the
amplitude $A$ of the soliton is only a function of $P$, $\rho$, and
$n$, it is also independent of time. $H( P,n,\rho) $   is obtained from the
exact Hamiltonian by substituting our variational ansatz (\ref{var1})
into (\ref{H}) .
It is convenient to write the  collective coordinate Hamiltonian in the form
\begin{equation}
H(P,n,\rho)= P^{3/2} (H_1 + z H_2),   \hspace{.5in} z={\kappa \over \sqrt{P}},
\end{equation}
where
\begin{eqnarray}
H_1 &=& {1 \over 2n} ({2n \over F_1})^{3/2} F_2, \nonumber \\
H_2 &=&  { 1  \over F_1}
({1 \over 2 })^{1/2n}  \rho \Gamma(1/2n) ,  \nonumber \\
F_{1}(\rho , n) & = & ({1  \over 2})^ {1/2n} \rho  \Gamma\left(1/2n
\right)  + \rho^{-1}  n^2 ( 2 )^{1/2n} \Gamma(2-1/2n), \nonumber \\
F_{2}(\rho, n) & = & ({1  \over 3})^ {1/2n} \rho \Gamma\left(1/2n
\right)  + \rho^{-1} {4 \over 9} n^2 ({ 3 )^{1/2n} \Gamma(2-1/2n)
\label{F}.
\end{eqnarray}
In terms of the conserved momentum $P$, the amplitude $A$
 of the soliton  is given by:
\begin{equation}
A^2 = {2 P n \over F_1(\rho,n)}.
\end{equation}
{}From Hamilton's equations the velocity is
\begin{equation}
\dot{q} = {\partial H(P,n,\rho)  \over \partial P} = P^{1/2} ( {3 \over 2} H_1
+ z
H_2) \equiv  P^{1/2} F(z) .
\end{equation}
Thus we see immediately that when $\kappa=0$ and $P \neq 0$,
\begin{equation}
\dot{q} =  {3 \over 2} { H(P,n,\rho) \over P}.
\end{equation}

Extremizing the action with respect to the variational parameters
$\rho$ and $n$ is the same as extremizing the
collective coordinate  Hamiltonian with respect to these parameters.
   At fixed
momentum $P$, $H(P,n,\rho) $ depends only on $\rho, n, z$; hence
at the extremum:
\begin{equation}
{ \partial \bar{H} \over \partial \rho} = { \partial \bar{H}  \over
\partial n } = 0, \hspace{.2in} \bar{H} \equiv {H(P,n,\rho)\over P^{3/2}},
\label{var}
\end{equation}
 and $n$ and $ \rho$ are related  to  z :
\begin{equation}
\rho = \rho(z), \hspace{.2in}  n=n(z).
\end{equation}
In general the dependence on z has to be determined numerically by
solving (\ref{var}). For small z we will obtain a simple
analytic
expression for all the variational parameters, the velocity and the
energy in terms of z. .

First let us study in detail the case $\kappa=0$.
For this case the collective coordinate Hamiltonian is:
\begin{equation}
H(P,n, \rho) = (2 n)^{1/2} P^{3/2} F_2(\rho,n) F_1^{-3/2} (\rho, n) .
\end{equation}

We need to extremize this Hamiltonian with respect to $\rho$ and
$n$  sequentially.
First, we extremize with respect to $\rho$. From (\ref{F}) we write
\begin{eqnarray}
F_1&=& a_1 \rho + b_1 / \rho, \nonumber \\
F_2&=& a_2 \rho + b_2 / \rho ,
\end{eqnarray}

where $a_1,a_2,b_1,b_2$ are known functions of $n$.
We then obtain
\begin{equation}
\rho^2(n)  = \left( -c_3 - \sqrt{c_3^2 - 4 c_1 c_2} \right) / 2c_1
\end{equation}
where $c_1 = - a_1 a_2$, $c_2 = b_1 b_2$, $c_3= 5 (a_2 b_1-a_1
b_2)$.
Extremizing the resulting Hamiltonian
$H_1= H_1(n,\rho(n))$   with respect to $n$ gives
\begin{equation}
n=1/2 , \hspace{.5in} \rho =1.
\end{equation}

We thus obtain the exact single peakon solution \cite{CH}  from
our variational approach:

\begin{equation}
u(x,t) = P^{1/2} e^{- |x- P^{1/2} t - x_0| }
\end{equation}
as well as the exact relationship: $ E = {2 \over 3} P^{3/2}$.

In the small z regime we can solve the minimization equations
(\ref{var}) analytically. We find

\begin{equation}
n \cong 1/2 + .924 z  \hspace{.5in} \rho \cong 1+ 2.498 z,  \label{n}
\end{equation}
which implies
\begin{equation}
\bar{H} (z) \cong (2/3) + 0.5 z ,    \hspace{.2in} F (z) \cong 1+ 0.5 z.
\label{lin}
\end{equation}

Equations (\ref{n}) and (\ref{lin}) allow us to
analytically determine the approximate soliton wave function for small
values of the parameter $ z$, since they determine $A, \beta, n,q(t)$.
For arbitrary $z$ the minimization is done numerically. Results for $z\leq 0.5$
are shown in Table 1 and are compared with the linear approximation.

\begin{table}[ht]
\caption{Full variational and linearized variational
solutions}
\centering
\begin{tabular}{c|ll|ll|ll|c} \hline\hline
$z$ & $n_{var}$ & $n_{lin}$ & $\rho_{var}$ & $\rho_{lin}$ &
$F_{var}$
& $F_{lin}$ & $[F_{var} - A_{var}/P^{1/2}] $\\
\hline
0 & 0.5 & 0.5 & 1.0 & 1.0 & 1.0 & 1.0 & 0.0  \\
0.1 & 0.589 & 0.592 & 1.259 & 1.250 & 1.054 & 1.050 & 0.080
\\
0.2 & 0.648 & 0.685 & 1.484 & 1.500 & 1.115 & 1.100 & 0.180
\\
0.3 & 0.688 & 0.777 & 1.683 & 1.749 & 1.182 & 1.150 & 0.282
\\
0.4 & 0.717 & 0.869 & 1.866 & 1.999 & 1.253 & 1.200 & 0.384
\\
0.5 & 0.739 & 0.962 & 2.038 & 2.249 & 1.327 & 1.250 & 0.486
\\
\hline\hline
\end{tabular}
\end{table}

\section{Exact scaling laws}

Let us now prove that the scaling result, which states that the
variational soliton amplitude  and velocity can be parametrized
 by a power of the conserved momentum, $P$, times a function of z,
 is also true for the exact solitons of the Camassa-Holm
equation (\ref{eq1}). Inserting a traveling wave solution, $f(\xi)$=$f(x-ct)$,
into  (\ref{eq1}) and integrating twice with the soliton boundary conditions,
 $f=f^{\prime}=f^{\prime\prime}=0$ at infinity,  we obtain:
\begin{equation}
{f^{\prime 2} \over  f^2} = {c- \kappa-f \over c-f} \label{eq:deriv}
\end{equation}
The maximum of the soliton occurs when $f^{\prime} =0 $ or when
\begin{equation}
f=c-\kappa
\end{equation}
This is the exact value for the amplitude $A$ of the soliton.
 Thus, if our variational solution were exact, we would find
\begin{equation}
F( z) - A/P^{1/2} = z.
\end{equation}
In the last column of Table 1 we display the variational result
 for the LHS of this equation (to be compared to $z$ in the first column).

One can write the conserved momentum $P$ for the soliton solution
as follows:
\begin{eqnarray}
P&=&{1 \over 2} \int_{-\infty} ^{\infty} d\xi \, (f^2 + f^{\prime 2}) \nonumber
\\
&=&\int_{0} ^{\infty} d\xi \,  f^2 \, \left[ 1+ {c- \kappa -f \over
c-f} \right].
\end{eqnarray}
On the half line we can change integration variables from $\xi$ to $f$
using (\ref{eq:deriv}) :
\begin{equation}
{d\xi \over df} = - { 1\over f} \left[ {c-f \over c- \kappa -f} \right]^{1/2}.
\end{equation}.

Therefore, we obtain
\begin{eqnarray}
P&=& \int_0^{c-\kappa} df \, \{ f \left[{c-f \over c-\kappa-f }\right]  ^{1/2}
+ f \left[{c- \kappa - f \over c-\kappa }\right]  ^{1/2} \} \nonumber \\
&=& c^2 \{ (1-{y \over 2}) \sqrt{1-y} - {y^2 \over 2} {\rm{log}} \left[y^{-1/2}
\, (1+\sqrt{1-y}) \right] \},
\end{eqnarray}
where
$y \equiv (\kappa/c)$.
Thus
\begin{equation}
P=c^2 g(y).
\end{equation}
 Since $ y= P^{1/2} {z \over c}$, we have $P^{1/2}/c = [g(z P^{1/2}/c)]^{1/2}$.
Thus we obtain the exact result that
\begin{equation}
\dot{q} = c = P^{1/2} F(z),
\end{equation}
where $F(z)$ is known implicitly from $g(y)$.
 For small z one can show that
\begin{equation}
c= P^{1/2} ( 1+ {z \over 2} + ...) ,
\end{equation}
so that our variational result gives the exact answer for small z.
Since the maximum amplitude $A$ of the soliton is
$A= c-\kappa$,  we find
\begin{equation}
A/ P^{1/2} = F(z) - z.
\end{equation}

Hence the scaling variable $z$ that we identified from the variational
method does indeed describe the soliton behavior at fixed momentum $P$.

\section*{Acknowledgements}
This work was supported in part by the DOE. F.C. would like to
thank Darryl Holm and Roberto Camassa  for  reading this paper,
 making valuable suggestions, and
for sharing their insights into this problem.

\end{document}